\documentclass[12pt]{article}
\usepackage[dvips]{graphicx}
\usepackage{pdproc}

  \makeatletter 
  \def\@cite#1{[#1]} 
  \makeatother    
\begin{document}

\renewcommand{\thefootnote}{\alph{footnote}}
\newcommand{\gsim}{ \mathop{}_{\textstyle \sim}^{\textstyle >} }
\newcommand{\lsim}{ \mathop{}_{\textstyle \sim}^{\textstyle <} }
\newcommand{\sla}[1]{\not\!#1}
\title{
Flavor and CP violation in the SUSY SU(5) GUT\\ with the right-handed
neutrinos }

\author{Junji Hisano}

\address{ 
ICRR, University of Tokyo\\
5-1-5 Kashiwa-no-Ha\\
Kashiwa City, 277-8582, Japan 
\\ {\rm E-mail: hisano@icrr.u-tokyo.ac.jp}}

\abstract{
The neutrino flavor mixing may induce rich flavor and CP violating
phenomena in the SUSY GUTs.  In this paper we discuss the hadronic
EDMs and the correlation among low-energy processes induced by the
mixings between second and third generations in the SUSY SU(5) GUT
with the right-handed neutrinos.  }

\normalsize\baselineskip=15pt

\section{Introduction}

The supersymmetric grand unified models (SUSY GUTs) predict rich
flavor and CP violation in the SUSY breaking terms for squarks and
sleptons, when origin of the SUSY breaking is  dynamics above the
GUT scale, such as the gravity mediation \cite{Hall:1985dx}.  This may
give a chance to probe the interactions at the GUT scale by the
low-energy flavor and CP violating processes, such as
$K^0$--$\overline{K}^0$ mixing, $B$ physics, and lepton flavor
violation, and EDMs. The recent results for CP asymmetries in $b$--$s$
penguin processes, including $B\rightarrow
\phi K_s$, by the Belle and Babar experiments \cite{belle,babar} are 2.4 and 2.7 $\sigma$
deviated from the Standard Model (SM) prediction, respectively. The
deviation might come from the flavor violation induced by the
GUT-scale interactions.

While the flavor and CP violating phenomena at low energy are indirect
probes to the SUSY GUTs, it is important to study the characteristic
signatures. One of them is the correlation between hadronic and
leptonic processes. In the SUSY GUTs quarks and leptons are embedded
into the common multiplets, and the SUSY breaking terms for the
squarks and sleptons is related to each others. For example, the
$b$--$s$ transition in the right-handed current is correlated with the
$Br(\tau\rightarrow\mu\gamma)$ in the models. 

Another one is the leptonic and hadronic electric dipole moments
(EDMs)
\cite{Barbieri:1995tw,Romanino:1996cn,Hisano:2004pw}.
While the processes are flavor diagonal, they depend on the flavor
violation in the internal sfermion lines in the SUSY
models. Especially, when both the left-handed and right-handed
sfermions have flavor violation, the EDMs are enhanced by the heavier
fermion masses. In the minimal SUSY standard model (MSSM), the sizable
flavor violating SUSY breaking terms for the left-handed squarks are
induced by the large top quark Yukawa coupling, and those for the
left-handed sleptons may be also generated by the neutrino Yukawa
interaction in the SUSY seesaw mechanism \cite{bm}. In the SUSY GUTs
the flavor violating SUSY breaking terms for the SU(5) partners of the
left-handed squarks or sleptons, right-handed squarks and sleptons,
are generated by the flavor violating interactions for the colored
Higgs multiplets, that are also SU(5) partners of the doublet Higgs
multiplets. Thus, non-vanishing EDMs are also a good signature of the
SUSY GUTs. 

In this article we discuss some flavor and CP violating phenomena in
the SUSY SU(5) GUT with the right-handed neutrinos. The neutrino
mixing generates new flavor and CP violation in squark and slepton
sectors. First, we discuss the hadronic EDMs. The neutrino Yukawa
coupling generates the flavor violating SUSY breaking terms for the
right-handed down-type squarks. This implies that we can investigate
the neutrino sector by the hadronic EDMs. Next, we show the
correlations among flavor and CP violating processes in the model,
assuming that the mixing between the second and third generations is
induced by the neutrino Yukawa coupling. The hadronic EDMs and
$Br(\tau\rightarrow\mu\gamma)$ are correlated with the CP asymmetry in
$b$--$s$ penguin processes, and they gives constraints on the deviation
from the SM predication. If one of them are discovered, others are good
tests for the model.

This article is organized as follows. We review the hadronic EDMs in
next section, and discuss the prediction for the hadronic EDMs in the
SUSY SU(5) GUT with the right-handed neutrinos in Section 3.  In
Section 4 we show the correlation among the hadronic EDMs, the CP
asymmetries in the $b$--$s$ penguin processes, and $Br(\tau
\rightarrow \mu \gamma)$. Section 5 is devoted to summary.

\section{Hadronic EDMs}

Now the most stringent bounds on the hadronic EDMs are those of
neutron \cite{Harris:jx} and $^{199}$Hg atom \cite{Romalis:2000mg},
and they give constraints on the SUSY models. In addition to them,
the improvement of the deuteron EDM is proposed recently
\cite{Semertzidis:2003iq}, and the sensitivity may reach to
$d_D\sim(1-3)\times 10^{-27}e\,cm$. If it is realized, one or two
orders of magnitude improvement may be achieved relative to the
current bounds on the CP violating parameters. In this section we
review the hadronic EDMs from the theoretical points of view.

The CP violation in the strong interaction of the light quarks is
dictated by the following effective operators,
\begin{eqnarray}
\label{CPV_quark}
{\cal L}_{\mathrm \sla{CP}} = 
 \overline{\theta}\, \frac{\alpha_s}{8\pi} G {\widetilde G}
+\sum_{q=u,d,s} i \frac{{\widetilde d}_q}{2} \overline{q}\, g_s(G\sigma)\gamma_5 q,
\label{eff1}
\end{eqnarray}
up to the dimension five ones. Here, $G_{\mu\nu}$ is the SU(3)
gauge field strength, ${\widetilde G}_{\mu\nu}=\frac{1}{2}
\epsilon_{\mu\nu\rho\sigma}G^{\rho\sigma}$ and 
$G\sigma=G^a_{\mu\nu}\sigma^{\mu\nu}T^a$.  The first term in
Eq.~(\ref{eff1}) is the effective QCD theta term, and the second term
is for the quark CEDMs.  The $\overline{\theta}$ parameter can be
$O(1)$ generically, however, it is strongly constrained by the neutron
EDM experiments. One of the most elegant solution is to introduce the
Peccei-Quinn (PQ) symmetry \cite{Peccei:1977hh}, since the axion makes
$\overline{\theta}$ vanish dynamically. On the other hand, when the
quark CEDMs are non-vanishing, the theta term is induced again even if
the PQ symmetry is introduced \cite{Bigi}.  Thus, in the evaluation of the effect
of the quark CEDMs, we need to include the QCD theta term
systematically. In the following we show the results in a case
that  the PQ symmetry is imposed.

We need to translate the quark-level interactions into the hadronic
interactions in order to calculate the hadronic EDMs.  This is a
rather difficult task, and the sizable theoretical uncertainties are
still expected for evaluation of hadronic matrix elements.  The PCAC
relation and the SU(3) chiral Lagrangian technique are useful for the
purpose. However, we still need some matrix elements.  The matrix
elements for the scalar operators $\overline{q}q$ are determined by
the baryon mass splittings and the nucleon sigma term
\cite{Zhitnitsky:1996ng} as
\begin{eqnarray}
\langle p|\overline{u}u |p\rangle =3.5,~
\langle p|\overline{d}d |p\rangle =2.8,~
\langle p|\overline{s}s |p\rangle =1.4.
\end{eqnarray}
The non-negligible strange quark component implies that that strange CEDM
gives sizable contributes to the hadronic EDMs. 

For the evaluation of those of the dipole operators $\overline{q} g_s
(G\sigma)q$, we need theoretical consideration.  In
Ref.~\cite{Zhitnitsky:1996ng,Falk:1999tm} it is argued that saturation
with the lightest $0^{++}$ state leads to the following relation in
the QCD sum rule,
\begin{eqnarray}
\label{eq:pospelov}
 \langle B_a \left| \overline{q} g_s (G\sigma)q
\right|B_b\rangle \simeq \frac{5}{3} m_0^2
\langle B_a \left| \overline{q}q\right|B_b\rangle,
\end{eqnarray}
for each quark.  This relation is, off course, approximate one.  The
latest QCD sum analysis of the CP violating pion-nucleon coupling
\cite{Pospelov:2001ys} shows that the result by evaluation of the
leading-order OPE term is almost consistent with that of the
saturation with the lightest $0^{++}$ state.  When the next and
next-to-next leading order term are included, the couplings are
suppressed by a factor and a possibility of the accidental
cancellation in the isoscalar pion coupling still remains. In the
following we assume the saturation with the lightest $0^{++}$ state
while keeping the uncertainties in mind.

The neutron EDM is induced by the one-loop diagrams of charged mesons
(Fig.~1(b)) while the EDM of $^{199}$Hg atom comes from the isovector
channel in the meson exchange between nucleons (Fig.~1(a)). $^{199}$Hg
atom is a diamagnetic atom, in which electrons make a close shell.  In
such atoms, the atomic EDMs are primary sensitive to the CP violation
in nucleons and represented by the Shiff moments. The recent
evaluation of the Shiff momentum of $^{199}$Hg atom shows that the
isoscalar and isotensor channel contributions in the $\pi$ and
$\eta^0$ exchanges are suppressed \cite{Dmitriev:2003sc}. Thus, the
contribution of $d_s^C$ to the EDM of $^{199}$Hg atom is generated by
the $\pi^0$--$\eta^0$.

\begin{figure}[htb]
\begin{center}
\includegraphics*[width=12cm]{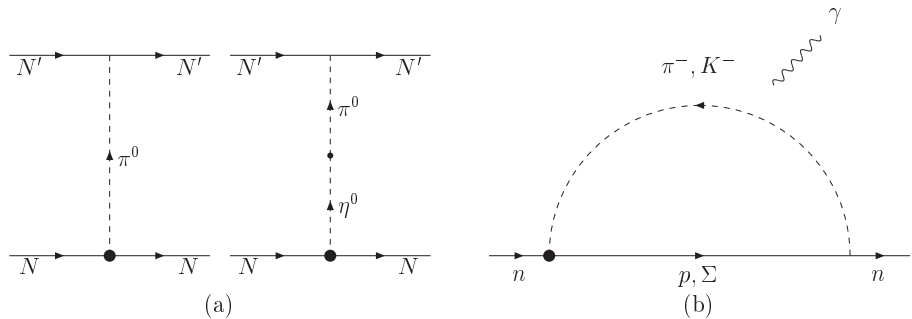}
\caption{%
a) Isovector contribution to the Shiff momentum of $^{199}$Hg atom. Here,
$N,N'=(p,n)$.
b) One-loop diagrams contributing to the neutron EDM. Large bobs represent
the CP violating nucleon-meson coupling.
}
\label{fig0}
\end{center}
\end{figure}

When the Peccei-Quinn symmetry is imposed, the EDMs for neutron and
$^{199}$Hg atom are given as \cite{HS2}
\begin{eqnarray}
d_n &=& -1.6 \times e (d_u^C+0.81\times d_d^C+0.16\times d_s^C),
\nonumber\\
d_{\rm Hg}&=&-8.7\times 10^{-3}\times e(d_u^C-d_d^C+0.005\times d_s^C),
\end{eqnarray}
respectively. Here, we do not include the contribution of the local
counter terms to the neutron EDM since they cannot be not fixed by
symmetry argument. The contribution of ${d}^C_s$ to the $^{199}$Hg
atom EDM, which is suppressed by the $\pi^0$--$\eta^0$ mixing
\cite{HS2}, is much smaller than that those of other light quarks. The
current experimental bounds on the EDMs for neutron
\cite{Harris:jx} and $^{199}$Hg atom \cite{Romalis:2000mg} are
\begin{eqnarray}
 |d_{n}| < 6.3 \times 10^{-26} e\, cm,
\nonumber\\
 |d_{\rm Hg}| < 1.9\times 10^{-28} e\, cm,
\label{dhbound}
\end{eqnarray}
respectively (90\%C.L.).
Thus, the upperbounds on the quark CEDMs are
\begin{eqnarray}
e|{d}^C_u|&<&3.9(2.2)\times 10^{-26}\;e\;cm, \nonumber\\
e|{d}^C_d|&<&4.8(2.2)\times 10^{-26}\;e\;cm,\nonumber\\
e|{d}^C_s|&<&2.4(44) \times 10^{-25}\;e\;cm,
\label{nedm_pq}
\end{eqnarray}
from the EDM of neutron ($^{199}$Hg atom). Here, we assume that the
accidental cancellation among the CEDMs does not suppress the EDMs.
Notice that we do not include the theoretical uncertainties which
comes from the matrix elements here.

The deuteron EDM is represented as 
\begin{eqnarray}
d_D&=&(d_n+d_p)+d^{NN}_D,
\label{edmd}
\end{eqnarray}
where $d_p$ is the proton EDM and the second term comes from the CP
violating nuclear force.  The nuclear dynamics in deuteron is rather
transparent, and the theoretical uncertainty is expected to be small.
When the PQ symmetry is imposed, each components in Eq.~(\ref{edmd})
are given as \cite{HS2}
\begin{eqnarray}
(d_p+d_n) &=& 
(-5.1\times \tilde{d}_u
+2.1\times \tilde{d}_d
+0.32\times \tilde{d}_s)\;e\;cm, \nonumber\\
d^{\pi NN}_D&=&
(
-11\times \tilde{d}_u
+11\times \tilde{d}_d
-0.063\times \tilde{d}_s)\;e\;cm,
\label{edm_d}
\end{eqnarray}
respectively.  From these equations, the measurement of the deuteron
EDM will be a stringent test for the SM. If the
sensitivity of $d_D
\sim(1-3)\times 10^{-27}e\,cm$ is established, $e\tilde{d}_u\sim
e\tilde{d}_d\sim 10^{-28}\;e\;cm$ can be probed from the nuclear
force. Also, for the strange quark CEDM, $e\tilde{d}_s\sim
10^{-26}\;e\;cm$ may be possible from the nucleon EDMs.

\section{Hadronic EDMs in the SUSY SU(5) GUT with the right-handed neutrinos}

Let us discuss the hadronic EDMs in the SUSY SU(5) GUT with the
right-handed neutrinos.  First, we review the flavor structure in the
squark and slepton mass matrices in the SUSY SU(5) GUT with the
right-handed neutrinos. The Yukawa interactions for quarks and leptons
and the Majorana mass terms for the right-handed neutrinos in this
model are given by the following superpotential,
\begin{eqnarray}
W&=& 
\frac14 f_{ij}^{u} \Psi_i \Psi_j H 
+\sqrt{2} f_{ij}^{d} \Psi_i \Phi_j \overline{H}
+f_{ij}^{\nu} \Phi_i \overline{N}_j {H}
+M_{ij} \overline{N}_i \overline{N}_j,
\label{superp_gut}
\end{eqnarray}
where $\Psi$ and $\Phi$ are for {\bf 10}- and {$\bf \bar{5}$}-dimensional
multiplets, respectively, and $\overline{N}$ is for the right-handed
neutrinos.  $H$ ($\overline{H}$) is {\bf 5}- ({$\bf \bar{5}$}-)
dimensional Higgs multiplets.  After removing the unphysical degrees
of freedom, the Yukawa coupling constants in Eq.~(\ref{superp_gut})
are given as follows,
\begin{eqnarray}
f^u_{ij} &=& 
V_{ki} f_{u_k} {\rm e}^{i \varphi_{u_k}}V_{kj}, \nonumber\\
f^d_{ij} &=& f_{d_i} \delta_{ij},\nonumber\\
f^\nu_{ij} &=& {\rm e}^{i \varphi_{d_i}} 
U^\star_{ij} f_{\nu_j}.
\label{yukawa}
\end{eqnarray}
Here, $\varphi_{u}$ and $\varphi_{d}$ are CP violating phases inherent
in the SUSY SU(5) GUT. They satisfy $\sum_i \varphi_{f_i} =0$
$(f=u$ and $d)$.  The unitary matrix $V$ is the CKM matrix in the extension
of the SM to the SUSY SU(5) GUT, and each unitary matrices $U$ and $V$
have only a phase. When the Majorana mass matrix for the right-handed
neutrinos is diagonal in the basis of Eq.~(\ref{yukawa}), $U$ is the MNS matrix
observed in the neutrino oscillation.  In this paper we assume the
diagonal Majorana mass matrix in order to avoid the complexity due to
the structure. In this case the light neutrino mass eigenvalues are
given as
$m_{\nu_{i}}
=
{f_{\nu_i}^2}\langle H_f \rangle^2/{M_{N_i}} $
where $H_f$ is a doublet Higgs in $H$. 

The colored Higgs multiplets $H_c$ and $\overline{H}_c$ are introduced
in $H$ and $\overline{H}$ as SU(5) partners of the Higgs doublets in
the MSSM, respectively. They have new flavor violating interactions in
Eq.~(\ref{superp_gut}). If the SUSY-breaking terms in the MSSM are
generated by dynamics above the colored Higgs masses, such as in
the gravity mediation, the sfermion mass terms may get sizable
corrections by the colored Higgs interactions. The interactions are
also baryon-number violating, and then proton decay induced by the
colored Higgs exchange is a serious problem, especially in the minimal
SUSY SU(5) GUT \cite{dim5}. However, it depends on the detailed
structure in the Higgs sector \cite{dim5sup,dim5sup6}. Thus, we ignore
the proton decay while we adopt the minimal Yukawa structure in
Eq.~(\ref{superp_gut}).

In the minimal supergravity scenario the SUSY breaking terms are
supposed to be given at the reduced Planck mass scale ($M_G$). In this
case, the flavor violating SUSY breaking mass terms at low energy are
induced by the radiative correction, and they  are qualitatively given
in a flavor basis as
\begin{eqnarray}
(m_{{\tilde{u}_L}}^2)_{ij}  &\simeq&-
V_{i3}V_{j3}^\star  \frac{f_{b}^2}{(4\pi)^2}
\;\; (3m_0^2+ A_0^2) \;\; 
(2 \log\frac{M_G^2}{M_{H_c}^2}+ \log\frac{M_{H_c}^2}{M_{SUSY^2}}),\nonumber\\
(m_{\tilde{u}_R}^2)_{ij}  &\simeq& -
{\rm e}^{-i\varphi_{u_{ij}}} 
V_{i3}^\star V_{j3} \frac{2f_{b}^2}{(4\pi)^2}
\;\; (3m_0^2+ A_0^2) \;\; 
\log\frac{M_G^2}{M_{H_c}^2}, \nonumber\\
(m_{{\tilde{d}_L}}^2)_{ij}  &\simeq&-
V_{3i}^\star
V_{3j} \frac{f_{t}^2}{(4\pi)^2} 
\;\; (3m_0^2+ A_0^2) \;\; 
(3 \log\frac{M_G^2}{M_{H_c}^2}+ \log\frac{M_{H_c}^2}{M_{SUSY}^2}),\nonumber\\
(m_{\tilde{d}_R}^2)_{ij}  &\simeq&-
{\rm e}^{i\varphi_{d_{ij}}}  U^\star_{ik}U_{jk} 
\frac{f_{\nu_k}^2}{(4\pi)^2} 
\;\; (3m_0^2+A_0^2) \;\; 
\log\frac{M_G^2}{M_{H_c}^2},\nonumber\\
(m_{\tilde{l}_L}^2)_{ij}  &\simeq&-
U_{ik}U_{jk}^\star
\frac{f^2_{\nu_k} }{(4\pi)^2} 
\;\; (3m_0^2+ A_0^2) \;\; 
\log\frac{M_G^2}{M_{N_k}^2},\nonumber\\
(m_{{\tilde{e}_R}}^2)_{ij}  &\simeq&-
{\rm e}^{i\varphi_{d_{ij}}} 
V_{3i}V^\star_{3j}
\frac{3 f_{t}^2}{(4\pi)^2} 
\;\; (3m_0^2+ A_0^2)\;\; 
\log\frac{M_G^2}{M_{H_c}^2},
\label{sfermionmass}
\end{eqnarray}
with $i\ne j$, where
$\varphi_{u_{ij}}\equiv\varphi_{u_{i}}-\varphi_{u_{j}}$ and
$\varphi_{d_{ij}}\equiv\varphi_{d_{i}}-\varphi_{d_{i}}$ and
$M_{H_c}$ is the colored Higgs mass.  Here,
$M_{SUSY}$, $m_0$ and $A_0$ are the SUSY-breaking scale in the MSSM
and the universal scalar mass and trilinear coupling,
respectively. $f_t$ is the top quark Yukawa coupling constant  while $f_b$ is
for the bottom quark. As mentioned above, the off-diagonal components in the
right-handed squarks and slepton mass matrices are induced by the
colored Higgs interactions, and they depend on the CP violating phases
in the SUSY SU(5) GUT with the right-handed neutrinos
\cite{Moroi:2000tk}.

When both the left-handed and right-handed squarks have the
off-diagonal components in the mass matrices, the EDMs and CEDMs for
the light quarks are enhanced significantly by the heavier quark mass.
The CEDMs of the down-type quarks are generated by the diagrams in
Fig.~2. In the SUSY SU(5) GUT with the right-handed neutrinos, the
neutrino Yukawa coupling induces the flavor violating mass terms for
the right-handed down-type squarks.  Since the flavor violating mass
terms for the left-handed down-type squarks are expected to be
dominated by the radiative correction induced by the top quark Yukawa
coupling as in Eq.~(\ref{sfermionmass}), we can investigate or
constrain the structure in the neutrino sector.

\begin{figure}[htb]
\begin{center}
\includegraphics*[width=7cm]{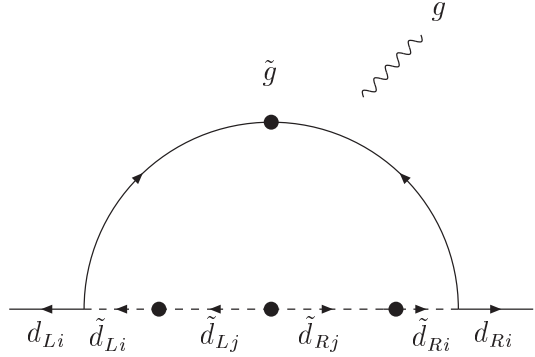}
\caption{%
Dominant diagram contributing to the CEDMs of down and strange
quarks when both the left-handed and right-handed squarks have 
flavor mixings.
}
\label{fig1}
\end{center}
\end{figure}

The CEDMs of the down and strange quarks derived by the flavor
violation in the both the left-handed and right-handed quark mass
matrices are given by the following dominant contribution, which is
enhanced by the heavier quark masses,
\begin{eqnarray}
 {d}^C_{d_i} = c\frac{\alpha_s}{4\pi}\frac{m_{\tilde g}}{\overline{m}^2_{\tilde d}}
f\left(\frac{m_{\tilde{g}}^2}{\overline{m}^2_{\tilde{d}}}
\right)  {\mathrm Im}
\left[(\delta^d_{ij})_{L}(\delta^d_{j})_{LR}(\delta^d_{ji})_{R}
\right],
\label{SUSYEDM}
\end{eqnarray}
where $m_{\tilde{g}}$ and $\overline{m}_{\tilde{d}}$ are the gluino and averaged
squark masses and $c$ is the QCD correction, $c\sim 0.9$.  The mass
insertion parameters are defined as
\begin{eqnarray}
(\delta^d_{ij})_{L/R}\equiv
\frac{(m_{{\tilde{d}_{L/R}}}^2)_{ij}}{(\overline{m}_{{\tilde{d}}}^2)},
&&
( \delta_{i}^{d})_{LR} 
\equiv  \frac{m_{d_i}(A_i^{(d)} -\mu\tan\beta)}{\overline{m}^2_{\tilde{d}}},
\end{eqnarray}
The function $f(x)$ is given as
\begin{eqnarray}
f(x)&=&\frac{177+118x-288x^2-6x^3-x^4+(54+300x+126x^2)\log x}{18(1-x)^6},
\label{mass_func}
\end{eqnarray}
and $f(1) =-11/180$. 

In Fig.~3 we show the CEDMs for the down and strange quarks in the
SUSY SU(5) GUT with the right-handed neutrinos. The figures come from
Ref.~\cite{Hisano:2004pw}.  In Fig.~3(a) the strange quark CEDM is
shown as a function of the right-handed tau neutrino mass.  We take
$M_{H_c}=2\times 10^{16}$GeV, $m_{\nu_\tau}=0.05$eV and
$U_{\mu3}=1/\sqrt{2}$.  For the SUSY breaking parameters to be fixed,
the minimal supergravity scenario is assumed, and $m_0=500$GeV,
$A_0=0$, $m_{\tilde{g}}=500$GeV and $\tan\beta=10$, which lead to
$\overline{m}_{\tilde{q}}\simeq 640$GeV. While the electron and muon
neutrino Yukawa interactions contribute to the flavor violation in the
right-handed down-type squark mass matrix, they are ignored here. They are
bounded by the constraints from the $K^0$--$\overline{K}^0$ mixing and
$Br(\mu\rightarrow e\gamma)$ when $|U_{e2}|\sim 1/\sqrt{2}$.  From
this figure, the right-handed tau neutrino mass should be smaller than
$\sim 3\times 10^{14}$GeV.

In Fig.~3(b) the down quark CEDM is shown as a function of the
right-handed tau neutrino mass. This comes from non-vanishing $U_{e3}$
in our assumption that the right-handed neutrino mass matrix is
diagonal. The current bound is not significant even when $U_{e3}=0.2$.

The new technique for the measurement of the deuteron EDM has a great
impact on the quark CEDMs as mentioned in the previous section. The
sensitivity of $d_D
\sim 10^{-27}e\,cm$ may imply that we can probe the structure in the
neutrino sector even if $M_{N_3}\sim 10^{13}$GeV or $U_{e3} \sim 0.02$
\cite{Hisano:2004pw}.

\begin{figure}[htb]
\begin{center}
\includegraphics*[width=7cm]{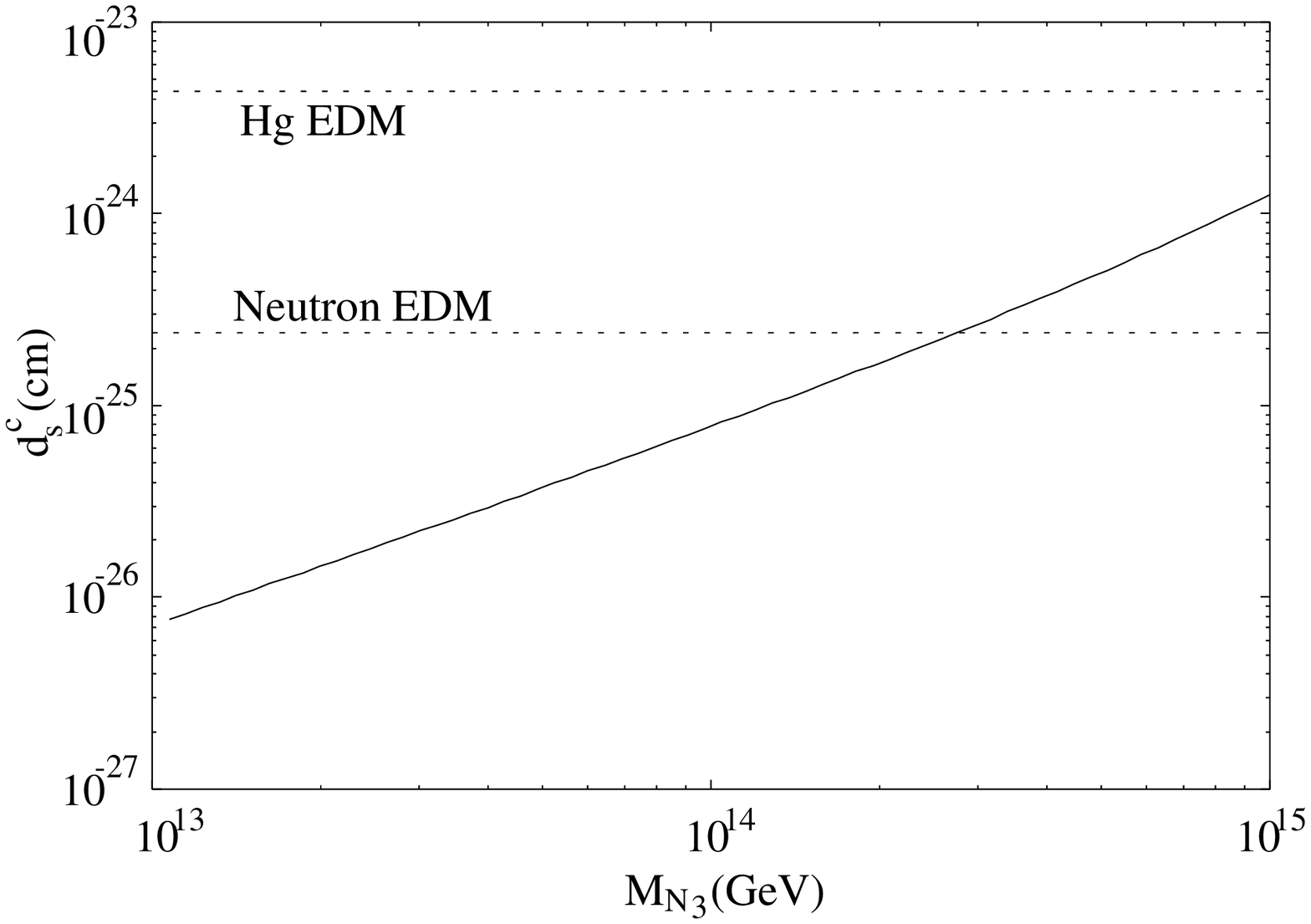}
\includegraphics*[width=7cm]{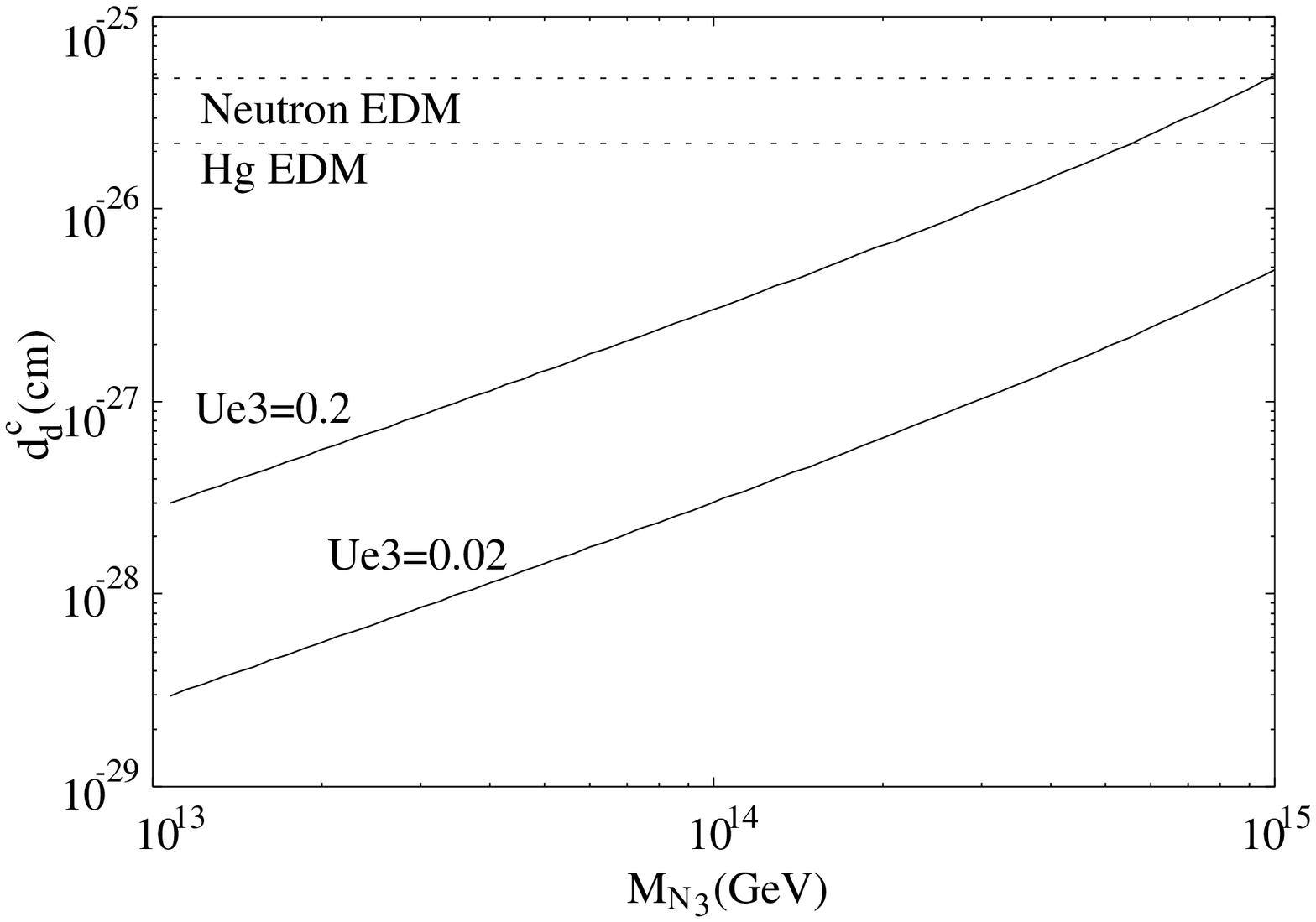}
\caption{%
CEDMs for the strange quark in (a) and for the down quark in (b) as
functions of the right-handed tau neutrino mass, $M_{N_3}$.  Here,
$M_{H_c}=2\times 10^{16}$GeV, $m_{\nu_\tau}=0.05$eV,
$U_{\mu3}=1/\sqrt{2}$, and $U_{e3}=0.2$ and 0.02.  For the MSSM
parameters, we take $m_0=500$GeV, $A_0=0$, $m_{\tilde{g}}=500$GeV and
$\tan\beta=10$. }
\label{fig2}
\end{center}
\end{figure}

We have not discussed the up quark CEDM in the model. The right-handed
up-type squarks also have the flavor violating mass terms, which
depend on the GUT CP violating phases, and the magnitudes are
controlled by the CKM matrix at the GUT scale. (See
Eq.~(\ref{sfermionmass}).)  They also contribute to the hadronic EDMs
in the SUSY SU(5) GUT
\cite{Romanino:1996cn}.  However, the off-diagonal terms in both the
left-handed and right-handed up-type squark mass matrices are induced
by the bottom quark Yukawa coupling, and the up quark CEDM and EDM
induced by them are proportional to $\tan^4\beta$. It is found that
the CEDM for the up quark can reach to $10^{-28}cm$ when $\tan\beta
\simeq 35$ and $m_{SUSY}\simeq 500$GeV. Thus, if we observe the
non-vanishing deuteron EDM larger than $10^{-27}e cm$, it might be
interpreted as the contribution of the down or strange quark CEDM in
the SUSY GUT with the right-handed neutrinos.

\section{Correlation among processes induced by mixings between the second and third generations}

The flavor and CP violating processes predicted in the SUSY GUTs
are indirect probes. Thus, it is important to take correlations among
various processes. The recent results for CP asymmetries in $b$--$s$
penguin processes, such as $B\rightarrow \phi K_s$, by the Belle and
Babar experiments may be explained by introduction of the flavor
violation between the second and third generations in the right-handed
down-type squark mass matrix, which is suggested by the atmospheric
neutrino oscillation. The $b$--$s$ penguin processes are sensitive to
the models beyond the SM \cite{jpsiphi}, and the various studies are
done for the processes in the SUSY models after the anomaly is
reported
\cite{iroiro}.  In this section we discuss the correlation among the
CP asymmetry in $B\rightarrow \phi K_s$ ($S_{\phi K_s}$),
$Br(\tau\rightarrow\mu\gamma)$, and the hadronic EDMs in the SUSY
SU(5) GUT with the right-handed neutrinos. Other processes induced by
the mixing between the second and third generations are given in
Ref.~\cite{Hisano:2003bd}.

First, we show the correlation between ${\tilde d}_s$ and $S_{\phi
K_s}$. When $(\delta_{32}^{(d)})_{R}$ is not vanishing, $S_{\phi
K_s}$ may have a sizable deviation from the SM prediction. The dominant
contribution to $S_{\phi K_s}$ is supplied by a penguin diagram with
the double mass insertion of $(\delta_{32}^{(d)})_{R}$ and
$(\delta_{3}^{(d)})_{LR}$ in Fig.4~(b). The contribution is
represented as $ H= - C_8^{R} {g_s}/({8\pi^2})
m_b\overline{s_R}(G\sigma) b_L$, where
\begin{eqnarray}
 C_8^{R}&=&\frac{\pi \alpha_s}{m^2_{\tilde{q}}}
\frac{m_{\tilde{g}}}{m_b}
(\delta_{3}^{(d)})_{LR}(\delta_{32}^{(d)})_{R}~
g\left(\frac{m_{\tilde{g}}^2}{\overline{m}^2_{\tilde{d}}}
\right),
\label{c8ap}
\end{eqnarray}
up to the QCD correction. Here,
\begin{eqnarray}
 g(x)&=&-\frac{53 - 9x - 45 x^2 + x^3 + 
    ( 18 + 66x + 12x^2 )\log x}{6\,
    {( 1-x ) }^5}
\end{eqnarray}
In a limit of $x\rightarrow 1$, $g(x)$ is 7/60.  On the other hand,
since the left-handed down-type squarks have the flavor violating mass
terms induced by the top quark Yukawa coupling, the strange quark CEDM
is also predicted by the diagram in Fig.~4(a).  From
Eqs. (\ref{SUSYEDM}) and (\ref{c8ap}), we find a strong correlation
between ${\widetilde d}_s$ and $C_8^R$ as \cite{HS}\cite{HS2}
\begin{eqnarray}
{\widetilde d}_s &=& -\frac{m_b}{4\pi^2} \frac{11}{21}
{\mathrm{Im}}\left[( \delta_{23}^{(d)})_{L} C_8^{R}\right]
\label{massin}
\end{eqnarray}
for $x=1$, up to the QCD correction. The coefficient $11/21$ in
Eq.~(\ref{massin}) changes from 1 to $1/3$ for $0<x<\infty$.

In Fig.~5, the correlation between $\widetilde{d}_s$ and $S_{\phi
K_s}$ is shown, assuming Eq.~(\ref{massin}) up to the QCD correction.
This figure comes from Ref.~\cite{HS2}.  In this figure, we take
$(\delta_{23}^{(d)})_{L} =-0.04$, ${\mathrm{arg}}[C_8^{R}]=\pi/2$ and
$|C_8^R|$ corresponding to
$10^{-5}<|(\delta_{32}^{(d)})_{R}|<0.5$. The matrix element of
chromomagnetic moment in $B\rightarrow \phi K_s$ is parameterized by
$\kappa$.  Since $\kappa$ may suffer from the large hadron
uncertainty, we show the results for $\kappa=-1$ and $-2$. From this
figure, the deviation of $S_{\phi K_s}$ from the SM prediction due to
the gluon penguin contribution should be suppressed when the
constraints on $\widetilde{d}_s$ from the hadronic EDMs, especially,
the neutron EDM, are applied.  Therefore, the hadronic EDMs give an
important implication to $S_{\phi K_s}$.

Here, we stress the theoretical uncertainties in the evaluation for
the hadronic EDMs, again. In this paper we assume the saturation with
the lightest $0^{++}$ state for the evaluation of the matrix element
$\langle B_a \left|
\overline{q} g_s (G\sigma)q \right|B_b\rangle$ as in Eq.~(\ref{eq:pospelov}), 
and our result relies on this assumption strongly. Also, the neutron
EDM evaluation does not include the local counter term
contribution. These uncertainties might still allow the sizable
deviation of $S_{\phi K_s}$ from the SM prediction. The further
experimental and theoretical improvement of the bounds on the hadronic
EDMs is very important.

\begin{figure}[htb]
\begin{center}
\includegraphics*[width=14cm]{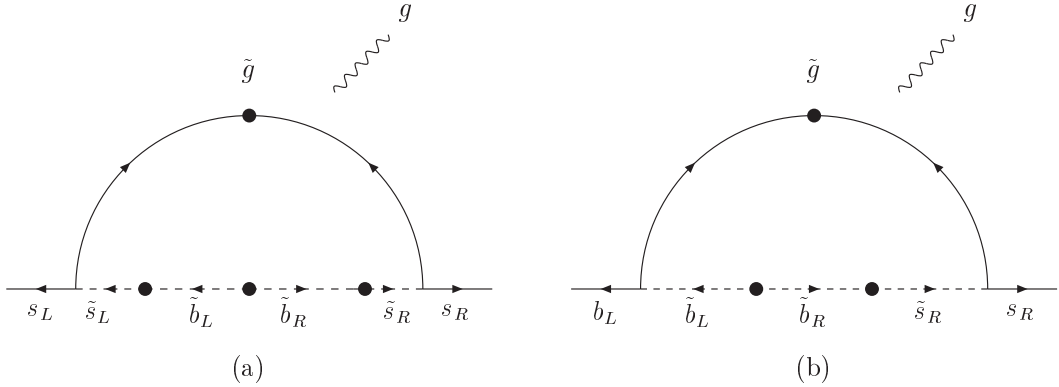}
\caption{%
(a) The dominant diagram contributing to the strange quark CEDM
 when both the left-handed and right-handed squarks have 
flavor mixings.
(b) The dominant SUSY diagram contributing to the CP asymmetry in
$B\rightarrow \phi K_s$ when the right-handed squarks have a mixing. 
}
\label{fig3}
\end{center}
\end{figure}

\begin{figure}[htb]
\begin{center}
\includegraphics*[width=10cm]{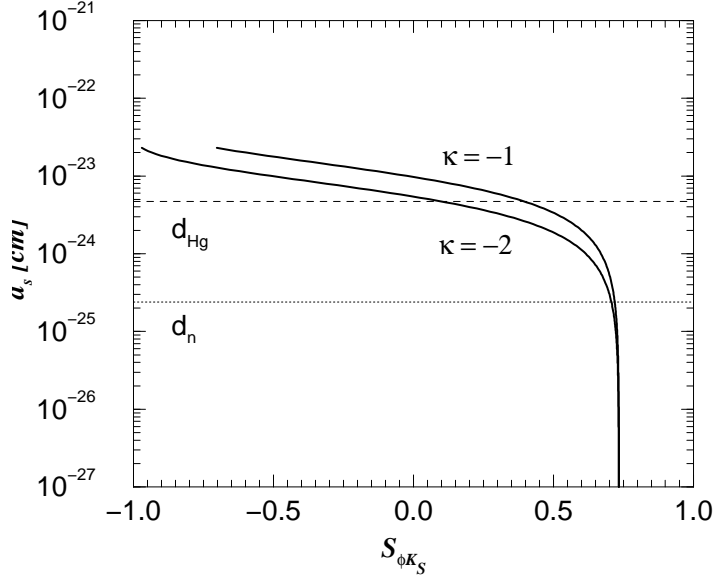}
\caption{%
The correlation between ${\widetilde d}_s$ and $S_{\phi K_s}$ assuming
${\widetilde d}_s = -{m_b}/(4\pi^2)(11/21)
{\mathrm{Im}}[ ( \delta_{23}^{(d)})_{L}C_8^{R}]$. Here, $( \delta_{23}^{(d)})_{L}
=-0.04$ and ${\mathrm{arg}}[C_8^{R}]=\pi/2$. $\kappa$ comes from 
the matrix element of chromomagnetic
moment  in $B\rightarrow \phi K_s$.
The dashed (dotted) line is the upperbound on ${\widetilde d}_s$ from the EDM of 
$^{199}$Hg atom (neutron). 
}
\label{fig4}
\end{center}
\end{figure}

Next, we show the correlation between $S_{\phi K_s}$ and
$Br(\tau\rightarrow\mu \gamma)$. The LFV processes are originally
predicted in the SUSY seesaw model \cite{bm,lfv}. As shown in
Eq.~(\ref{sfermionmass}) the (2,3) components in the left-handed
slepton and right-handed down-type squark mass matrix are related as
\cite{Hisano:2003bd}
\begin{eqnarray}
(m_{\tilde{d}_R}^2)_{23}^\star 
&\simeq&
{\rm e}^{-i\varphi_{d_{23}}}
\frac{\log\frac{M_G}{M_{H_c}}}{\log\frac{M_G}{M_{N_3}}}
(m_{\tilde{l}_L}^2)_{23}.
\end{eqnarray}
Thus, the observables are also correlated.  

\begin{figure}[htb]
\begin{center}
\includegraphics*[width=10cm]{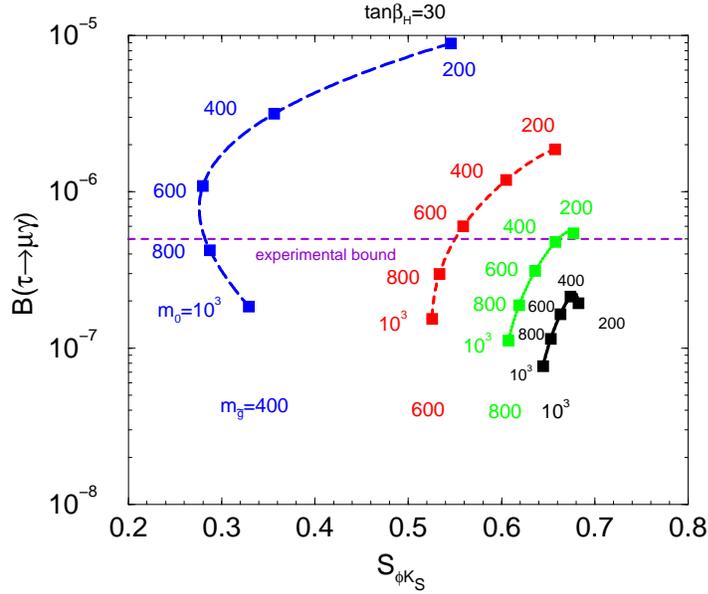}
\caption{%
$Br(\tau\rightarrow\mu\gamma)$ as a function of $S_{\phi K_S}$ for
fixed gluino masses $m_{\tilde{g}}=400$, 600, 800,
and 1000\,GeV. $\tan\beta$ is 30. Also,
200\,GeV$<m_0<$1\,TeV, $A_0=0$, $m_{\nu_\tau}=5\times10^{-2}$\,eV,
$M_{N}=5\times 10^{14}$\,GeV,
and $U_{32}=1/\sqrt{2}$. $(\varphi_{d_2}-\varphi_{d_3})$ is taken for the
deviation of $S_{\phi K_S}$ from the SM prediction to be maximum. The
constraints from $b\rightarrow s\gamma$ and the light-Higgs  mass
are imposed.}
\label{fig5}
\end{center}
\end{figure}

In Fig.~\ref{fig5} we show $Br(\tau\rightarrow\mu\gamma)$ as a
function of $S_{\phi K_S}$ for fixed gluino masses $m_{\tilde{g}}$ in
the SUSY SU(5) GUT with the right-handed neutrinos. 
This figure comes from Ref.~\cite{Hisano:2003bd}. Here, $\tan\beta$ is
30, 200\,GeV$<m_0<$1\,TeV, $A_0=0$,
$m_{\nu_\tau}=5\times10^{-2}$\,eV, $M_{N_3}=5\times 10^{14}$\,GeV, and
$U_{32}=1/\sqrt{2}$. The deviation of $S_{\phi K_s}$ is maximized when
$m_{\tilde{g}}$ is lighter and $m_0$ is comparable to $m_{\tilde{g}}$.
While larger $\tan\beta$ enhances the deviation of $S_{\phi K_s}$ from
the SM prediction ($\sim 0.7$), it is also bounded by the constraint
from $Br(\tau\rightarrow\mu \gamma)$. If the results of the Belle and 
Babar experiments are confirmed in the future, The search for 
$\tau\rightarrow\mu \gamma$ is also a good check for the SUSY SU(5) GUT
with the right-handed neutrinos.

\section{Summary}

While the low-energy flavor and CP violating observables are predicted
in the SUSY GUTs, they are the indirect probes. Thus, it is important
to study the characteristic signatures. One way is to check the
correlation among various processes. Especially, hadronic and leptonic
processes are related to each others due to the GUT relation. Second
is the EDMs. They are induced when both mixings of the left-handed and
right-handed sfermions are non-vanishing.  In this paper we discuss
the hadronic EDMs and the correlation among low-energy processes
induced by mixing between the second and third generations in the SUSY
SU(5) GUT with the right-handed neutrinos.

\bibliographystyle{plain}

\end{document}